# Understanding Dynamics in Coarse-Grained Models: IV. Connection of Fine-Grained and Coarse-Grained Dynamics with the Stokes-Einstein and Stokes-Einstein-Debye Relations

Jaehyeok Jin[1,2] and Gregory A. Voth[1,*]

[1] Department of Chemistry, Chicago Center for Theoretical Chemistry, Institute for Biophysical Dynamics, and James Franck Institute, The University of Chicago, Chicago, IL 60637, USA

[2] Department of Chemistry, Columbia University, New York, NY 10027, USA

* Corresponding author: gavoth@uchicago.edu

**Abstract**

Applying an excess entropy scaling formalism to the coarse-grained (CG) dynamics of liquids, we discovered that missing rotational motions during the CG process are responsible for artificially accelerated CG dynamics. In the context of the dynamic representability between the fine-grained (FG) and CG dynamics, this work introduces the well-known Stokes-Einstein and Stokes-Einstein-Debye relations to unravel the rotational dynamics underlying FG trajectories, thereby allowing for an indirect evaluation of the effective rotations based only on the translational information at the reduced CG resolution. Since the representability issue in CG modeling limits a direct evaluation of the shear stress appearing in the Stokes-Einstein and Stokes-Einstein-Debye relations, we introduce a translational relaxation time as a proxy to employ these relations, and we demonstrate that these relations hold for the ambient conditions studied in our series of work. Additional theoretical links to our previous work are also established. First, we demonstrate that the effective hard sphere radius determined by the classical perturbation theory can approximate the complex hydrodynamic radius value reasonably well. Also, we present a simple derivation of an excess entropy scaling relationship for viscosity by estimating the elliptical integral of molecules. In turn, since the translational and rotational motions at the FG level are correlated to each other, we conclude that the "entropy-free" CG diffusion only depends on the shape of the reference molecule. Our results and analyses impart an alternative way of recovering the FG diffusion from the CG description by coupling the translational and rotational motions at the hydrodynamic level.

# I. Introduction

The dynamics of molecular systems at the atomistic resolution are composed of various diffusive motions such as translation and rotation,[1-3] and it is of the utmost importance to characterize these motions in the liquid state.[4,5] In this light, numerous transport properties, e.g., the diffusion coefficient, shear viscosity, and structural relaxation times, have been extensively investigated by combining experimental and computational studies. However, relatively little attention has been given to the dynamics at reduced representations in multiscale modeling. In the process of developing a coarse-grained (CG) model by renormalizing the complex fine-grained (FG) degrees of freedom,[6-14] correct fluctuation and dissipation forces controlling the dynamical information at the reduced resolution can be rigorously described using the Mori-Zwanzig formalism.[15-18] However, inferring these fluctuation and dissipation interactions from the stochastic integro-differential equations are often computationally expensive and pose some numerical issues for complex systems.[19-24]

Alternatively, one can evolve the CG variables using only the Hamiltonian equation of motion, but the missing friction and fluctuations in the Hamiltonian mechanics often result in an accelerated CG dynamics.[19, 21, 25-28] Since the dissipative friction kernel appearing in the Mori-Zwanzig equation of motion is intrinsically a many-body quantity, an accurate estimation of this acceleration factor and its physical meaning has not been clearly elucidated.[29-33] This paper series (denoted as Papers I to III[34-36] according to the numbering in their titles) strives to resolve this discrepancy in the CG dynamics by providing a CG dynamic representability based on the alternative framework known as the excess entropy scaling relationship. As suggested by Rosenfeld in 1977,[37] this semi-empirical scaling relationship indicates that the reduced transport properties, e.g., diffusion coefficient or shear viscosity, is proportional to the system's reduced (molar) excess entropy. For example, the reduced self-diffusion coefficient $D^*$ is expressed as

$$D^* = D_0 \exp(\alpha s_{ex}), \qquad (1)$$

where the reduced (molar) excess entropy $s_{ex}$ is a measure of how much the system entropy $S$ deviates from the ideal gas value $S_{id}$ at the same thermodynamic state points $\rho$ and $T$, as given by

$$s_{ex} = \frac{S_{ex}}{Nk_B} = \frac{1}{Nk_B}\big(S(\rho,T) - S_{id}(\rho,T)\big). \qquad (2)$$

Even though this quasi-universal scaling relationship is not rigorously derived from first-principle physics (note that several attempts have been made to partially derive this relationship[38-42]), we demonstrated in Paper I[34] that this scaling relationship holds its universality for the same molecular entities in the FG and CG representations, extending the applicability of the Rosenfeld scaling to CG systems. Furthermore, the semi-empirical nature of the relationship was addressed for the CG system by developing a statistical mechanical theory to understand the CG diffusion as a hard sphere diffusion process, which is extensively discussed in Paper II.[35]

Subsequently, Paper III[36] focused on a more fundamental question arising in CG dynamics: why are CG dynamics under the Hamiltonian mechanics faster than the FG counterpart? We ascribed this to the missing motions during the CG process. For example, in the single-site CG model, the rotational motions are lost at the CG resolution, and thus the resultant CG model does not have any rotational diffusion. Furthermore, there will not be any momentum exchange between the different motions upon the collision of the CG molecules, highlighting that there is also no translation-rotation coupling at the CG model level, which usually slows down the collective diffusion processes. Based on these observations, we developed a systematic procedure to restore the FG dynamics information to the CG model. Briefly stated, we first extracted the rotational dynamics information from the FG system and incorporated it into the CG system. Then, upon integrating the effective rotational dynamics, we evaluated the translation-rotation coupling parameter[43-48] based on the non-sphericity of the molecule[49, 50] to account for the momentum changes in angular and linear momentum upon collision. In turn, our designed approach was able to recover the reference FG diffusion coefficients and provide the correct time scale relative to the CG diffusion.[36] Yet, there was one critical limitation: prior knowledge of the rotational diffusion at the FG level is required in order to correct the CG dynamics. This implies that the aforementioned approach might face a circular argument because we do not have information about the rotational diffusion of the target system during the CG process. We also note that the bottom-up CG parametrization such as force-matching[51-55] or relative entropy minimization[56-58] only requires statistical information of the reduced configurations to determine the CG interactions, in which only the FG translational motions are left at this reduced resolution, and the motions beneath the CG resolution are integrated out.

This work derived from spatial scales at opposites extremes is designed to resolve the above issues by providing a clearer understanding of the translational and rotational diffusive motions at an atomistic level. At the macroscopic hydrodynamics level, it is widely recognized that the translational diffusion of large particles (solute) immersed in fluids (solvents) with much smaller size follows what is known as the Einstein relation,[59] expressed as:

$$D_{\text{trans}} = \frac{k_B T}{\zeta},$$

(3)

where $k_B$ is the Boltzmann constant, $T$ is the temperature, and $\zeta$ denotes the effective friction constant of the solute particle.[60] To further determine the friction coefficient, when a solute is assumed to be spherical with radius $R$ under low-Reynolds number flow (non-turbulent),[61] the Stokes law[62] can be applied to estimate the friction constant:

$$\zeta = c\pi\eta R,$$

(4)

where $\eta$ is the (shear) viscosity of the neat solvent, and $c$ is a constant dependent on the hydrodynamic boundary conditions applied at the solute surface. Specifically, $c$ equals 4 for the "slip" condition, implying zero normal component of solvent velocity at the interface, while $c = 6$ is associated with the "stick" condition, where the velocity of the solvent on the solute surface equals that of the solute (same relative velocity).[63] Due to these boundary conditions, the exact form of $\zeta$ and $D_{\text{trans}}$ varies depending on the system. By combining kinetic theory [Eq. (3)] and continuum hydrodynamics [Eq. (4)], the Stokes-Einstein (SE) relation can be derived as

$$D_{\text{trans}} = \frac{k_B T}{6\pi\eta R}.$$

(5)

Analogous to $D_{\text{trans}}$, the rotational hydrodynamic motion leads to the rotational diffusion coefficient of a single (colloidal) solute:

$$D_{\text{rot}} = \frac{k_B T}{\zeta^r},\tag{6}$$

where $\zeta^r$ represents the rotational friction drag coefficient of a solute sphere. Under the same assumption in deriving the SE relation, the Stokes-Einstein-Debye (SED) relation[64] can be derived that links $D_{\text{rot}}$ to $\eta$ and $R$:

$$D_{\text{rot}} = \frac{k_B T}{f 8\pi \eta R^3},\tag{7}$$

where $f$ is the hydrodynamic boundary condition factor for rotations that equals 1 under the stick boundary condition and decays to 0 for the complete slip boundary condition. For non-stick and non-spherical colloids, several studies have characterized the non-trivial $f$ factor as well.[65-68] The theoretical elegance and simplicity of both the SE and SED relations have proven useful for almost a century, with successful applications in elucidating the dynamics of numerous chemical and physical systems.[69]

More importantly, while Eqs. (3)-(7) are derived and rooted in the continuum-level description under the critical assumption that the size of the diffusing particle is significantly larger than the solvent molecules, it has been experimentally found that extending the SE/SED relations to tracer molecules that are similar in size to the solvent molecules at the molecular level is valid.[70-77] In such molecular liquids, the solute size $R$ should be replaced with the so-called *hydrodynamic radius* $R_h$, and the diffusion coefficients become self-diffusion coefficients. Consequently, computer simulations of liquids have focused on determining what extent the SE and SED relations hold by investigating the decoupling between molecular diffusion and shear viscosity.[78-82] As the SE and SED relations are intrinsically coupled to each other, our central argument is based on the idea that the rotational diffusion might be inferred from the translational diffusion, which is different from what we found in Paper III.[36] Therefore, this paper aims to combine the SE and SED relations with the excess entropy scaling formalism in order to retrieve the missing rotational information upon the coarse-graining process based on the reduced CG configurations. Having both translational and rotational dynamics from the SE and SED relations, we seek to provide an alternative interpretation of the CG dynamics by presenting the dynamic representability (or "dynamic correspondence") between the FG and CG dynamics.

The remainder of this paper is organized as follows. In Sec. II, we briefly review Papers I-III[34-36] and introduce computational approaches to employ the SE and SED relations in CG systems. Next, we check the validity of the SE and SED relations in Sec. III. We derive two different equations: (1) the reduced scaling relationship from the SE and SED relations and (2) a new expression for the CG diffusion based on the SE and SED relations. Finally, concluding remarks are given in Sec. IV.

## II. Theory
### A. Review: Excess Entropy Scaling for FG and CG Systems
Before establishing a link between the SE/SED relations and excess entropy relationships, some notation and earlier findings discussed in the preceding papers will be reviewed first.

In Paper I, we have introduced Eqs. (1) and (2) to the molecular systems at two different resolutions: FG and CG.[34] The reduced diffusion coefficients for both resolutions are scaled by the same macroscopic units given by Newtonian mechanics as[83, 84]

$$D^* = D \frac{\rho^{\frac{1}{3}}}{\left(\frac{k_B T}{m}\right)^{\frac{1}{2}}}, \tag{8}$$

where $\rho$ is the system number density ($N/V$) with the molecule mass $m$. Therefore, based on Eq. (8), we attributed the differences in the FG and CG diffusion coefficients to the following two factors: the changes in the excess entropy term $\exp(\alpha s_{ex})$ and the $D_0$ value in Eq. (1) denoted as the "entropy-free" diffusion coefficient. This immediately suggests that the Rosenfeld scaling could be applied to understand the CG dynamics with respect to the reference FG diffusion process, but establishing a one-to-one correspondence was highly ambiguous until Papers I[34] and II[35] due to two reasons. First, there was no guarantee if both the FG and CG systems follow an identical scaling [i.e., same $\alpha$ from Eq. (1)]. Also, no physical understanding of $D_0$ was available because of its semi-quantitative yet empirical nature.[85] Based on our recent understanding of the differences in the FG and CG entropies using the entropy-enthalpy decomposition,[86] Paper I[25] proposed a new framework by combining the two-phase thermodynamic method with the systematic theories given by Lazaridis, Karplus,[87] and Zielkiewicz.[88] Utilizing this framework, we further demonstrated that the FG and its corresponding CG systems undergo the same scaling relationship, extending the applicability of the Rosenfeld relationship. For water, these two scaling relationships were obtained as (from Paper I[34]):

$$\ln D^*_{FG} = \ln D_0^{FG} + \alpha^{FG} s_{ex}^{FG} = 0.73 \times s_{ex}^{FG} + 2.15, \tag{9}$$

$$\ln D^*_{CG} = \ln D_0^{CG} + \alpha^{CG} s_{ex}^{CG} = 0.70 \times s_{ex}^{CG} - 0.35. \tag{10}$$

Even though both Eqs. (9) and (10) were scaled with $\alpha^{FG} \approx \alpha^{CG} = 0.7$, the artificially accelerated CG dynamics can only be partially explained. Because of a relatively larger excess entropy term $s_{ex}^{CG} > s_{ex}^{FG}$, the correspondence between $D^*_{FG}$ and $D^*_{CG}$ is still not clear unless differences in $D_0$ terms are resolved.

To clarify this ambiguity, in Paper II, we developed a statistical mechanical theory to analytically derive the $D_0$ term for the CG liquid system.[35] Since the single-site CG system exhibits only translational motions, we mapped the CG system into an effective hard sphere system conserving the dynamical properties by capturing the long-wavelength fluctuations from the system

$$S(k = 0)_{CG} = S(k = 0)_{HS}, \tag{11}$$

where $S(k)$ is the structure factor computed using the wave vector $k$, and the limit $k = 0$ on the left-hand side gives $S(k = 0) = 1 + 4\pi\rho \int_0^\infty dr\, R^2[g(R) - 1]$ that can be determined from the CG simulation. On the other hand, the right-hand side of Eq. (11) can be expressed as

$$S(k = 0)_{HS} = \rho k_B T \kappa_T. \tag{12}$$

Since the isothermal compressibility $\kappa_T = -(\partial V/\partial P)_T/V$ is a function of the packing density $\varphi$ for hard spheres, solving Eq. (11) yields a unique $\varphi$ value that can conserve the CG dynamics.

Then, $\varphi$ can be used to determine $D_0^{CG}$ because the hard sphere dynamics can be expressed as an analytical function of the packing density using kinetic theory.[89, 90] We have tested various equations of state used to describe the hard sphere systems and found that the $D_0^{CG}$ values for water at different temperatures were well reproduced regardless of the relative complexity or accuracy of the chosen equations of state. For example, by employing the Carnahan-Starling equation of state,[91] the $D_0^{CG}$ can be derived as

$$D_0^{CG}(\varphi) = \frac{\pi^{1/6}}{48} 6^{4/3} \frac{(1-\varphi)^3}{\varphi^{2/3}(2-\varphi)} \exp\left[\frac{4\varphi - 3\varphi^2}{(1-\varphi)^2}\right]. \tag{13}$$

**B. Review: Role of Rotational Diffusion**
Having established a physical understanding of $D_0^{CG}$, our subsequent interest was to elucidate the differences between $D_0^{FG}$ and $D_0^{CG}$, which was the main scope of Paper III.[36] As discussed in Sec. I of this paper, the main difference between the FG and CG diffusion processes is that the relatively finer motions beneath the CG resolution are lost during the CG process. That being said, for single-site CG models, we envisaged that these missing rotational motions should be responsible for the changes in $D_0^{CG}$ upon the coarse-graining process.

Since the excess entropy scaling relationship [Eq. (1)] is designed for translational diffusion, we obtained the effectively projected translational displacements from the rotational motions based on our findings in Paper II that the CG liquids (especially water) can be viewed as hard sphere entities with an effective hard sphere radius of $R_{HS}$,[35]

$$\vec{R}_{t-rot} \cong R_{HS} \cdot \vec{\phi}_I(t). \tag{14}$$

Interestingly, we further corroborated that the resultant effective rotational diffusion, $D_{t-rot}$, follows the universal scaling relationship derived from the translational motion,

$$D^*_{t-rot} = D_0^{t-rot} \exp(\alpha^{t-rot} s_{ex}^{t-rot}), \tag{15}$$

where $\alpha^{t-rot} = \alpha^{CG} \approx \alpha^{FG}$, indicating the high fidelity of our projection method. Finally, when plugging the "entropy-free" diffusion coefficient from the rotational motions $D_0^{t-rot}$ back to the CG model, we simultaneously introduced the translational-rotational coupling, which accounts for the angular and linear momentum changes upon collision.[43-48] Chandler suggested that once the translational-rotational coupling is in place, the realistic diffusion coefficient from rough hard spheres $D_{RHS}$ is smaller than that of smooth hard spheres $D_{SHS}$[92]

$$D_{RHS} = A_D D_{SHS}, \tag{16}$$

where the roughness parameter $A_D$ is a measure of the translational-rotational coupling ranging from zero to unity (i.e., perfectly smooth hard sphere).[92-96] Even though the roughness parameter quantifies how close the system is to a smooth sphere, in which momentum exchange does not occur upon collision, it might be an obscure and rather intractable quantity by definition. This ambiguity can be clarified by adopting an alternative approach from Ruckenstein and Liu,[50] where they sought to understand the roughness parameter as a function of the intrinsic molecular properties. This approach was possible by examining the acentric factor $\omega$,[97, 98] originally introduced by Pitzer to describe the non-sphericity of liquids by quantifying the deviations in the behavior of non-ideal liquids from the theorem of corresponding states.[49] From 42 different data sources, Ruckenstein and Liu found the following relationship between $A_D$ and $\omega$,

$$A_D = 0.9673 - 0.2527\omega - 0.70\omega^2. \tag{17}$$

Combining Eqs. (14)-(17), we estimated the roughness parameter for water from its acentric factor, and recovered the rotational diffusion to the CG representation in the presence of the translational-rotational coupling

$$\ln D_0^{FG} \approx \ln D_0^{CG} + \ln D_0^{t-rot} + \ln A_D, \tag{18}$$

where our central assumption is that the translational and rotational degrees of freedom are decoupled. This assumption is analogous to the decoupling of the translational and rotational structure factors of water in quasi-elastic neutron scattering studies.[99-101] In order to indirectly estimate $D_{rot}$, for the remainder of Sec. II, we introduce an alternative computational protocol to employ the SE/SED relations for our system.

## C. Viscosity and Representability Issue

In order to determine the translational and rotational diffusion coefficients using the SE and SED relations, one must consider a shared variable between the SE/SED relations: the shear viscosity $\eta$. In general, in a similar approach for calculating the diffusion coefficient,[2] $\eta$ is often obtained using the Green-Kubo formalism

$$\eta = \frac{V}{k_B T} \cdot \int_0^\infty dt \langle \mathcal{P}_{\alpha\beta}(0) \cdot \mathcal{P}_{\alpha\beta}(t) \rangle, \tag{19}$$

where the stress tensor element $\mathcal{P}_{\alpha\beta}$ can be chosen from the five independent shear components of the stress tensor: $\mathcal{P}_{xy}, \mathcal{P}_{xz}, \mathcal{P}_{yz}, (\mathcal{P}_{xx} - \mathcal{P}_{yy})/2$, and $(\mathcal{P}_{yy} - \mathcal{P}_{zz})/2$.[102] For example, without loss of generality, an *xy*-direction will give a tensor element $\mathcal{P}_{xy}$ of the form

$$\mathcal{P}_{xy} = \frac{1}{V} \sum_I \left[ m_I v_I^x v_I^y + \sum_{J>I} (R_I^x - R_J^x) F_{IJ}^y \right], \tag{20}$$

where $v_I^x$ (or $v_I^y$) is the *x*- (or *y*-) component of the velocity vector for the CG site *I*, $(R_I^x - R_J^x)$ denotes the *x*-component of the displacement vector linking the CG sites *I* and *J*, and $F_{IJ}^y$ is the *y*-component of the effective force acting on the CG site *I* from the CG site *J*. Even though Eq. (20) is defined in the CG representation, the "actual" shear viscosity can be computed from the atomistic trajectories using the atomistic virial by summing over each atom assigned to the CG representation.[103, 104]

However, after propagating the CG model, an exact determination of the CG virial from the CG trajectory is much more difficult due to the following two issues. The first issue is from the sampling problem that occurs when computing both the FG and CG shear viscosities. It is well known that an accurate calculation of shear viscosity at many different temperatures and force fields is computationally challenging. In our case, we need to carry out 40 independent computations for shear viscosity, which may suffer from sampling issues despite the use of statistical techniques that can enhance the sampling of the pressure tensor using the isotropy of the system suggested by Daivis and Evans,[105, 106] or the time decomposition method.[107] Alternatively, this sampling problem could be lessened by employing different microscopic expressions, e.g., the Born-Green formula[108] or viscoelastic theory.[109]

Aside from the sampling issue, a more fundamental problem arises from the representability problem in bottom-up CG modeling.[12, 86, 110, 111] The representability issue states that conventional observable expressions defined at the atomistic resolution generally change in their CG counterpart due to the renormalization process, resulting in a lack of expressiveness of the approximate CG model. In particular, because of these missing degrees of freedom, naïvely estimated CG virials would be always smaller than the FG reference, resulting in an underestimation of the pressure tensors.[111] Furthermore, naïve estimation of virials is known to give non-physical CG pressure with negative values.[112] In order to accurately capture the CG virial, one needs to introduce new basis sets that effectively describe the CG interactions and are compatible with the virial expressions. These basis sets should be parametrized with respect to the FG information, as extensively illustrated in Ref. 112. Therefore, in this paper, we opt not to explicitly calculate the stress tensor and shear modulus.

**D**. **Structural Relaxation Time: Proxy to Viscosity**
Instead of directly calculating viscosity, an alternative approach that has been widely adopted in the previous literature is to utilize the *relaxation time.* In Sec. II D and Sec. II E, we provide a physical link between the viscosity and relaxation time.

While there are many different timescales corresponding to the various time correlations, a straightforward relaxation time related to viscosity is the Maxwell viscoelastic relaxation time[113] $\tau_M$ that is often obtained from the relaxation behavior of the autocorrelation function of the stress tensor components using a stretched exponential decay:

$$\frac{V}{k_B T} \langle \mathcal{P}_{\alpha\beta}(0) \cdot \mathcal{P}_{\alpha\beta}(t) \rangle = G_\infty \exp\left[\left(-\frac{t}{\tau_M}\right)^\gamma\right],$$

(21)

where $G_\infty$ is the infinite frequency shear modulus of the liquid. To note, $\gamma$ should be 1 for the Maxwell viscoelastic model,[114] which is consistent with observations from molecular simulations of water.[115] Hence, the Maxwell viscoelastic relaxation time is directly related to the viscosity of liquids and can be computed via

$$\tau_M = \frac{\eta}{G_\infty} = \int_0^\infty dt \frac{\sum_{\alpha\beta} \langle \mathcal{P}_{\alpha\beta}(0) \mathcal{P}_{\alpha\beta}(t) \rangle}{\sum_{\alpha\beta} \langle \mathcal{P}_{\alpha\beta}(0) \mathcal{P}_{\alpha\beta}(0) \rangle},$$

(22)

though computing $\tau_M$ from the Green-Kubo formalism[116] [right-hand side of Eq. (22)] still suffers from representability and sampling issues.

Given this context, a new relaxation time $\tau$, referred to as the structural relaxation time (or α-relaxation time) corresponding to the long-time regime of the translational dynamics, is often employed in place of $\tau_M$. Nevertheless, since there is no systematic relationship between $\tau_M$ and $\tau$, it is still unclear whether $\tau$ can be used instead of $\tau_M$ to account for viscosity. For example, Shi, Debenedetti, and Stillinger reported that structural and viscosity relaxation times were not equivalent at low temperatures in glass-forming systems.[117]

Notably, a study of the TIP4P/2005 model for water at different temperatures indicates that the relaxation time ratio $\tau_M/\tau$ remains nearly constant at ambient temperatures.[82] A noticeable change in $\tau_M/\tau$ only occurs below 285 K, indicating a decoupling between viscosity and relaxation processes. This observation is also consistent with the interpretation given by Ref. 118 in which the Maxwell viscoelastic relaxation timescale $\tau_M$ is in close agreement with the local atomic connectivity timescale $\tau_{LC}$ at temperatures higher than 285K. Since our system was prepared for ambient conditions ranging from 280K to 360K, there should not be any noticeable decoupling between the viscosity and relaxation processes, and thus it is reasonable to approximate the structural relaxation timescale $\tau_M$ as $\tau$ in our work.

### E. Structural Relaxation Time: Incoherent intermediate scattering function

In practice, the translational relaxation time $\tau$ is obtained from the incoherent intermediate scattering function, defined as

$$F_s(\mathbf{k}, t) = \langle \frac{1}{N} \sum_{I=1}^{N} \exp[i\mathbf{k} \cdot (\mathbf{R}_I(t) - \mathbf{R}_I(0))] \rangle,$$

(23)

where the configurational variable $\mathbf{R}_I(t)$ is the configuration of the CG site $I$ at time $t$. The wavenumber $\mathrm{k} = |\mathbf{k}|$ was chosen to be at the first peak position of the static structure factor $S(k)$ from the mapped CG (center-of-mass) configurations. The intermediate scattering function is the self-part of the Fourier transform of the Van Hove function, $G(\mathbf{r}, t)$, given as

$$F(\mathbf{k}, t) = \int d\mathbf{k}\, G(\mathbf{r}, t) e^{-i\mathbf{k}\mathbf{r}t}.$$

(24)

Note that $F(\mathbf{k}, t)$ is divided into two different contributions: the self $F_s(\mathbf{k}, t)$ contribution and the distinct $F_d(\mathbf{k}, t)$ contribution

$$F(\mathbf{k}, t) = \langle \frac{1}{N} \sum_{I=1}^{N} \sum_{J=1}^{N} \exp[i\mathbf{k} \cdot (\mathbf{R}_I(t) - \mathbf{R}_J(0))] \rangle = F_s(\mathbf{k}, t) + F_d(\mathbf{k}, t),$$

(25)

where $F_s(\mathbf{k}, t)$ is the contribution from $I = J$, and $F_d(\mathbf{k}, t)$ is otherwise ($I \neq J$). The self-part is called the incoherent intermediate scattering function, and it characterizes the relaxation time of the system, which can also be directly examined from inelastic neutron scattering experiments.

Unlike previous studies in the determination of the translational relaxation time of water, we use the reduced (CG) configurations, i.e., center-of-mass of the molecule, not the oxygen atom throughout this work. Even though we have demonstrated that the structural correlation functions based on the center-of-mass configurations do not noticeably deviate from the structural correlation functions of oxygen atoms,[119] we note that all the values and results may not be identical to that from the oxygen atoms. Since the reported structural relaxation times for water in similar conditions are in the order of 0.1–1 ps, we re-sampled the FG trajectories with more frequent sampling to collect the FG configurations during the simulations. Specifically, during the constant $NVT$ run using the Nosé-Hoover thermostat,[120, 121] we collected the FG configurations every 10 fs over the course of 1 ns. For each of the 100,000 frames from the FG trajectories, the incoherent intermediate scattering function was computed using the LiquidLib suite[122] with a frame interval of 10 fs, and the $F_s(k, t)$ was averaged every 0.1 ns.

After computing the $F(k, t)$ at the CG resolution, the long-time relaxation behavior (α-relaxation time) can be captured using the Kohlrausch-Williams-Watts (KWW) function

$$F_s(k, t) \approx (1 - f_c) \exp\left[-\left(\frac{t}{\tau_s}\right)^2\right] + f_c \exp\left[-\left(\frac{t}{\tau}\right)^{\beta_\alpha}\right],$$

(26)

Where the term $f_c \exp[-(t/\tau)^{\beta_\alpha}]$ describes the stretched exponential decaying with the degree of non-exponentiality $\beta_\alpha$.[123] The fitting procedure of the KWW function was carried out using MATLAB R2019b[124] with the trust-region-reflective least-squares regression.[125]

**F. CG Water Systems: BUMPer Model**

This series of papers is concerned with the molecular system of water. For the CG model, we specifically utilized a recently developed CG water model called the Bottom-up Many-Body Projected Water.[119, 126] The effective CG interaction underlying the BUMPer CG model has a pairwise form, but this effective pairwise interaction is obtained by projecting the three-body interaction of the Stillinger-Weber potential [Eq. (27)] form onto the pairwise basis sets:

$$U_{eff}^{(2)}(R_{IJ}) = \sum_{K>J} U^{(3)}(\theta_{JIK}, R_{IJ}, R_{IK})$$

$$= \sum_{K>J} \lambda_{JIK} \epsilon_{JIK} (\cos\theta_{JIK} - \cos\theta_0)^2 \exp\left(\frac{\gamma_{IJ}\sigma_{IJ}}{R_{IJ} - a_{IJ}\sigma_{IJ}}\right) \exp\left(\frac{\gamma_{IK}\sigma_{IK}}{R_{IK} - a_{IK}\sigma_{IK}}\right).$$

(27)

In practice, since our model is constructed via bottom-up, we first performed the force-matching from Multiscale Coarse-graining (MS-CG) methodology[51-55] to the center-of-mass mapped atomistic trajectories from the SPC/E,[127] SPC/Fw,[128] TIP4P/2005,[129] and TIP4P/Ice[130] force fields, resulting in the two-body $U_{3b}^{(2)}(R_{IJ})$ and three-body $U_{3b}^{(3)}(\theta_{JIK}, R_{IJ}, R_{IK})$ interaction terms:

$$U_{3b} = \sum_I \sum_{J>I} U_{3b}^{(2)}(R_{IJ}) + \sum_I \sum_{J \neq I} \sum_{K>J} U_{3b}^{(3)}(\theta_{JIK}, R_{IJ}, R_{IK}).$$

(28)

The three-body parametrization method was introduced by Ref. 131, and related discussions are followed in Ref. 132. Then, from Eq. (28), we extracted the effective conditional probability for triplet variables appearing in the Stillinger-Weber interaction[133] at the fixed pair distance, $p(\theta_{JIK}, R_{IK}|R_{IJ})$, and we numerically integrated the three-body interactions weighted by the conditional probability with respect to the auxiliary variables ($\theta_{JIK}$ and $R_{IK}$ in this case) such that

$$U_{eff}^{(2)}(R_{IJ}) = 2(N_c - 1) \int d\theta_{JIK} dR_{IK} p(\theta_{JIK}, R_{IK}|R_{IJ}) U^{(3)}(\theta_{JIK}, R_{IJ}, R_{IK}),$$

(29)

where the local coordination number $N_c$ appears in order to match the counting over triplets to the pair summation. The final BUMPer interaction is, thus, expressed as a summation of the two-body and effectively projected two-body interactions from the three-body interactions:

$$U_{3b} = \sum_I \sum_{J>I} U_{3b}^{(2)}(R_{IJ}) + \sum_I \sum_{J \neq I} \sum_{K>J} U_{3b}^{(3)}(\theta_{JIK}, R_{IJ}, R_{IK})$$
$$= \sum_I \sum_{J \neq I} \left\{ U_{3b}^{(2)}(R_{IJ}) + 2(N_c - 1) \int d\theta_{JIK} dR_{IK} p(\theta_{JIK}, R_{IK}|R_{IJ}) U_{3b}^{(3)}(\theta_{JIK}, R_{IJ}, R_{IK}) \right\}.$$

(30)

Details of the many-body projection theory and its performance are extensively discussed in Refs. 119, 126. Interestingly, in prior work, we have applied this new BUMPer model to low-temperature regimes to demonstrate that the model is capable of capturing the various hierarchical anomalies of water.[88] However, in this work, we are only interested in the ambient conditions where the temperature ranges from 280K to 360K corresponding to the liquid phase of water as the Rosenfeld scaling relationship is valid for ambient liquid phases. This condition removes the possibility of the SE and SED violations where it generally occurs for $T \leq 1.5 T_g$ where the glass transition temperature for water is $T_g \sim 136\,\text{K}$.[134-136] Even though this general rule of thumb coincides with the reported simulations of the SE and SED violations in water, a careful assessment of the validity of the SE and SED relations (especially at lower temperatures) is still needed.

**G. Computational Details: Translational and Rotational Diffusion**
In order to utilize the SE and SED relations, the translational and rotational (self) diffusion coefficients must be computed beforehand. We calculated these dynamical properties with different FG force fields for water at different temperatures in Paper III.[36] Here, we briefly describe the computational details we used to compute these properties.

Both the translational and rotational diffusion coefficients were computed using Einstein's relation. The translational diffusion coefficient is expressed using the center-of-mass mean square displacement (MSD) $\langle R^2(t) \rangle$ as

$$D_{\text{trans}} = \lim_{t \to \infty} \frac{1}{6t} \langle R^2(t) \rangle = \lim_{t \to \infty} \frac{1}{6t} \frac{1}{N_{CG}} \sum_{I}^{N_{CG}} |\vec{R}_I(t) - \vec{R}_I(0)|^2, \tag{31}$$

where $\vec{R}_I(t)$ denotes the configuration of the CG site $I$ at time $t$. Similarly, the rotational diffusion coefficient is defined using the rotational MSD $\langle \phi^2(t) \rangle$

$$D_{\text{rot}} = \lim_{t \to \infty} \frac{1}{4t} \langle \phi^2(t) \rangle = \lim_{t \to \infty} \frac{1}{4t} \frac{1}{N_{CG}} \sum_{I}^{N_{CG}} |\vec{\phi}_I(t) - \vec{\phi}_I(0)|^2, \tag{32}$$

where $\vec{\phi}_I(t)$ is a rotational displacement vector for the CG site $I$ at time $t$. Unlike the translational MSD [Eq. (31)], the rotational MSD [Eq. (32)] should be carefully calculated while ensuring it is unbounded. This is because $\vec{\phi}_I(t)$, by design, is bounded from 0 to $2\pi$. Following the procedure employed in Ref. 137, we instead constructed a normalized polarization vector $\hat{p}_I(t)$ to describe the differential of rotational displacement vectors. We define $\hat{p}_I(t)$ as the vector from the center-of-mass configuration of molecule $I$ [$\vec{R}_I(t)$] to the midpoint of two hydrogen atoms, $\vec{r}_{H1,I}$ and $\vec{r}_{H2,I}$

$$\hat{p}_I(t) := \frac{1}{2} (\vec{r}_{H1,I} + \vec{r}_{H2,I}) - \vec{R}_I(t). \tag{33}$$

Then, the differential of the rotational displacement from time from $t$ to $t + \delta t$ can be expressed as

$$|\Delta \vec{\phi}_I(t + \delta t)| := \cos^{-1}(\hat{p}_I(t) \cdot \hat{p}_I(t + \delta t)), \tag{34}$$

where the direction of $\Delta \vec{\phi}_I(t + \delta t)$ is determined by the cross product between $\hat{p}_I(t)$ and $\hat{p}_I(t + \delta t)$

$$\Delta \vec{\phi}_I(t + \delta t) \parallel \hat{p}_I(t) \times \hat{p}_I(t + \delta t). \tag{35}$$

Therefore, the finalized expression for the rotational displacement vector is given by numerically integrating $\Delta \vec{\phi}_I$ over the course of time

$$\vec{\phi}_I(t) \approx \sum_{i=t_0}^{t} \Delta \vec{\phi}_I \left( t + i \cdot \frac{t - t_0}{N_{\delta t}} \right). \tag{36}$$

Note that $\vec{\phi}_I(t)$ in Eq. (36) is no longer bounded because it does not depend on the size of the angular displacement, only on the differential.

It is worth noting that an alternative approach to extract the rotational dynamics is also possible based on the rotational correlation function,[138]

$$C_\ell(t) = \langle P_\ell[\hat{p}_I(t) \cdot \hat{p}_I(0)] \rangle, \tag{37}$$

where $P_\ell[\cdot]$ is the $\ell$th order Legendre polynomial. By fitting the computed $C_\ell(t)$ to the KWW-like function, another structure relaxation known as the rotational relaxation time $\tau_\ell$ can be inferred and can play a major role in the SE and SED relations. However, in this work, we choose to calculate the rotational diffusion coefficient based on Eq. (32), since it has been reported that the α-relaxation time $\tau$ is analogous to the higher-order rotational relaxation time.[139] In practice, we used the FG and CG trajectories generated from Paper III,[36] employing the publicly available software package, OpenMSCG.[140] The computational details regarding the initial setup and simulation details are thoroughly discussed in the aforementioned papers.

**H. SE/SED Relations for Water**

Two central factors should be clarified that influence the fidelity of the SE/SED relations for water systems before discussing the analyses of the molecular simulations and diffusion coefficient. First, while various theoretical studies[79-81, 139, 141-149] and experiments[75, 150-153] have reported violations of the SE/SED relations in liquids, these violations are typically observed in low-temperature regimes, e.g., supercooled states.[75-77, 154, 155] Therefore, given that our main focus is on ambient conditions of water, where the excess entropy scaling also holds, we expect the SE/SED relations will uphold without significant deviations.

Moreover, as outlined in the Introduction, the SE and SED relations are dependent on the hydrodynamic boundary conditions (slip or stick), which become less straightforward to evaluate and assess at the molecular level. For the translational diffusion, the boundary condition affects at most 33% of the diffusion, but this becomes particularly problematic when assessing the rotational diffusion as molecules deviate from stick conditions. Nevertheless, in this work, we assume that water under ambient conditions obey stick conditions for several reasons. By experimental study, Wilbur et al. characterized the ratio between $k_B T$ and $\pi R_h D \eta$ for water under different conditions, corresponding to the hydrodynamic boundary condition factor $c$ in Eq. (4).[156] The authors further observed stick-like behavior in $c$ factors, ranging from 4.9 to 6, at temperatures studied in this work. This experimental observation is supported by other computational studies that employ stick conditions to analyze the SE/SED breakdown for water at the atomistic level.[139, 157-160] At the CG level, there has been no systematic investigation into this relationship until this study. Nevertheless, we can reasonably estimate the boundary conditions based on the effective BUMPer interaction profile reported in Ref. 119. Schmidt and Skinner found in Ref. 161 that pair interactions with non-negligible attractions between non-sphere solute result in stick-like boundary conditions. This condition perfectly aligns with BUMPer at the single-site resolution, suggesting that the stick condition is plausible for both FG and CG level diffusion coefficients. Hence, throughout Sec. III, we specifically utilize the following SE/SED relations under stick boundary conditions:

$$D_\text{trans} = \frac{k_B T}{6\pi \eta R_h},$$

(38a)

$$D_\text{rot} = \frac{k_B T}{8\pi \eta R_h^3}.$$

(38b)

We note that additional improvement of Eqs. (38a) and (38b) is possible by deriving the hydrodynamic factor for translational diffusion

$$c = 6 \frac{1 + 4\eta/\beta R_h}{1 + 6\eta/\beta R_h},$$

(39)

where $\beta$ is the slip coefficient (or sliding friction) that exhibits the slip condition for $c$ when $\beta$ approaches 0 and stick when $\beta \to \infty$.[162, 163] Similarly, the hydrodynamic factor for the rotational diffusion can be considered based on the non-spherical geometry of the system.[65-68] Nevertheless, as the primary objective of this paper is to introduce the SE/SED relations to uncover the excess entropy scaling, pursuing this aspect is beyond the scope of the current work. Additionally, we will demonstrate later in Sec. III that the stick condition can effectively capture the structural and dynamical properties of water in quantitative manner. Therefore, we leave this area as a potential avenue for future study when applying the present framework to water or even more complex molecules.

## III. Results
### A. Diffusion Coefficients for Translational and Rotational Motions

We first calculated the translational and rotational diffusion coefficients of the FG water models using Eqs. (31) and (32), respectively. The corresponding diffusion coefficients at different temperatures are listed in Table 1.

**Table 1:** Translational and rotational diffusion coefficients of water evaluated for FG models at temperatures ranging from 280K to 360K at intervals of 20K: (a) SPC/Fw, (b) SPC/E, (c) TIP4P/2005, and (d) TIP4P/Ice.

| (a) SPC/E | | | (b) SPC/Fw | | |
|---|---|---|---|---|---|
| Temperature | $D_{\text{trans}}$ (Å$^2$·ps$^{-1}$) | $D_{\text{rot}}$ (rad$^2$·ps$^{-1}$) | Temperature | $D_{\text{trans}}$ (Å$^2$·ps$^{-1}$) | $D_{\text{rot}}$ (rad$^2$·ps$^{-1}$) |
| 280K | 1.5511×10$^{-1}$ | 7.6548×10$^{-2}$ | 280K | 2.1982×10$^{-1}$ | 1.0128×10$^{-1}$ |
| 300K | 2.5049×10$^{-1}$ | 1.0822×10$^{-1}$ | 300K | 3.5277×10$^{-1}$ | 1.6826×10$^{-1}$ |
| 320K | 3.9796×10$^{-1}$ | 1.5765×10$^{-1}$ | 320K | 4.8354×10$^{-1}$ | 2.1769×10$^{-1}$ |
| 340K | 5.3343×10$^{-1}$ | 2.0242×10$^{-1}$ | 340K | 6.0839×10$^{-1}$ | 2.7744×10$^{-1}$ |
| 360K | 6.1163×10$^{-1}$ | 2.6898×10$^{-1}$ | 360K | 6.7734×10$^{-1}$ | 3.3094×10$^{-1}$ |
| (c) TIP4P/2005 | | | (d) TIP4P/Ice | | |
| Temperature | $D_{\text{trans}}$ (Å$^2$·ps$^{-1}$) | $D_{\text{rot}}$ (rad$^2$·ps$^{-1}$) | Temperature | $D_{\text{trans}}$ (Å$^2$·ps$^{-1}$) | $D_{\text{rot}}$ (rad$^2$·ps$^{-1}$) |
| 280K | 1.2031×10$^{-1}$ | 5.9983×10$^{-2}$ | 280K | 6.0386×10$^{-2}$ | 3.2848×10$^{-2}$ |
| 300K | 2.1091×10$^{-1}$ | 1.0171×10$^{-1}$ | 300K | 1.2431×10$^{-1}$ | 5.7670×10$^{-2}$ |
| 320K | 3.2525×10$^{-1}$ | 1.5224×10$^{-1}$ | 320K | 1.8838×10$^{-1}$ | 9.3879×10$^{-2}$ |
| 340K | 4.4492×10$^{-1}$ | 2.0914×10$^{-1}$ | 340K | 2.9431×10$^{-1}$ | 1.4744×10$^{-1}$ |
| 360K | 6.2018×10$^{-1}$ | 2.7117×10$^{-1}$ | 360K | 3.9958×10$^{-1}$ | 2.0005×10$^{-1}$ |

In order to understand the relationship between $D_{\text{trans}}$ and $D_{\text{rot}}$ from Table 1, we computed the incoherent intermediate scattering function $F_s(\mathbf{k}, t)$ for each thermodynamic state point using Eq. (23), as shown in Fig. 1.

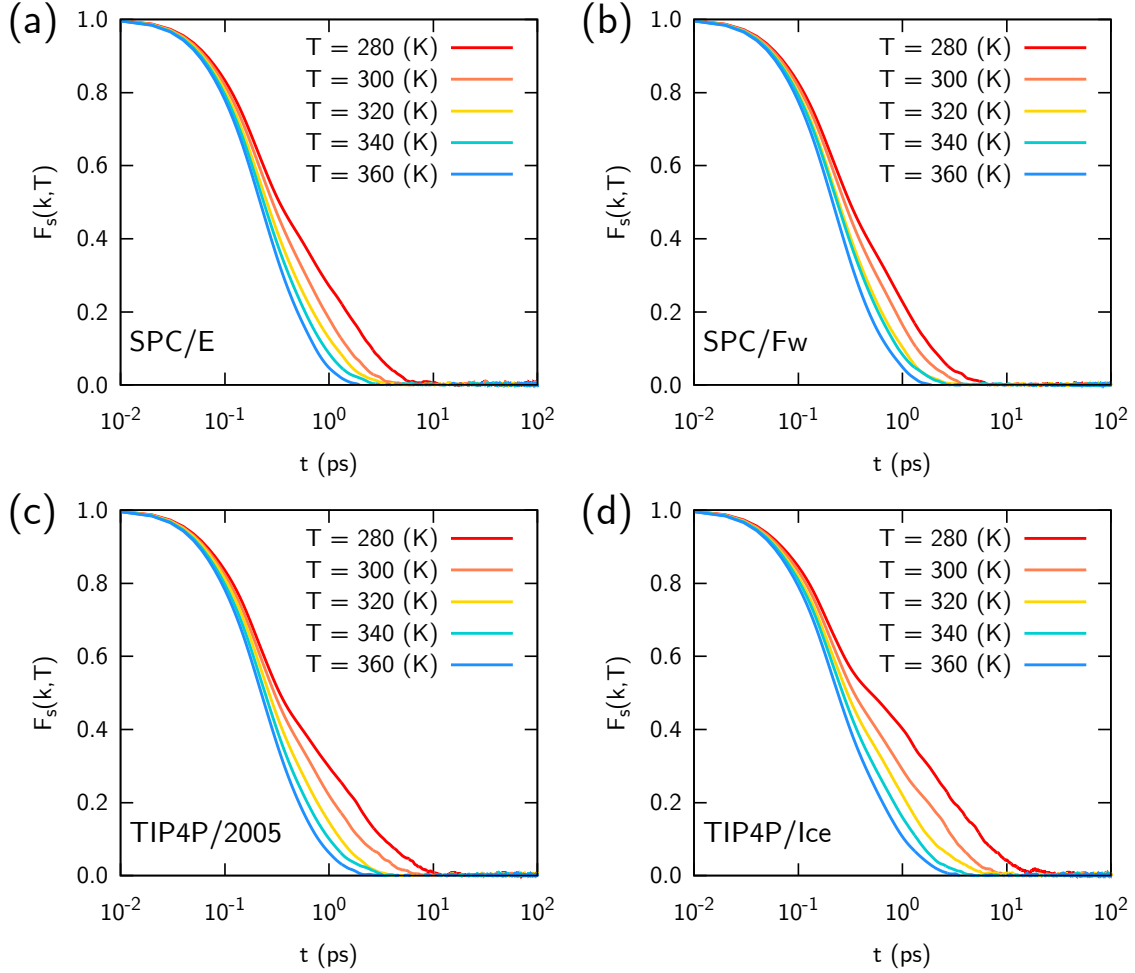

**Figure 1.** The incoherent intermediate scattering function $F_s(k, t)$ from the FG trajectories at temperatures ranging from 280K to 360K: (a) SPC/E, (b) SPC/Fw, (c) TIP4P/2005, and (d) TIP4P/Ice. From the computed $F_s(k, t)$, we fitted the KWW function [Eq. (26)] to capture the stretched exponential contribution $f_c \exp[-(t/\tau)^{\beta_\alpha}]$. Table S2 in the Supplementary Material lists the full results of parametrized variables.

The incoherent intermediate scattering function is known to exhibit one to two relaxation times depending on the temperature of the system. Regardless of the FG force fields, the relaxation time is longer at lower temperatures. Interestingly, we notice that the TIP4P/Ice force field has a pronounced decaying shoulder, which can be understood from the design principle of the TIP4P/2005 force field to match the freezing temperature of water and other associated behaviors at low temperatures.

The fitted relaxation times $\tau$ from Eq. (26) that are plotted against temperature in Fig. 2(a) range from 0.4 (higher temperature) to 1.5 ps (lower temperature), which is in good agreement with the reported timescales.[82, 117, 139, 148, 149, 164-168] Furthermore, we do not observe cases where the relaxation time becomes remarkably larger, indicating that we do not have to worry about supercooled-like dynamics for our cases, which can break down the SE or SED relation. A detailed analysis of the temperature dependence of $\tau$ will be provided in Sec. II. B, and a more straightforward validation of the SE and SED relations will be given in Sec II. C. We also computed $F_s(k, t)$ for the propagated CG trajectories to check if the spurious CG dynamics can be also represented by the incoherent intermediate scattering function. Figure S1 in the Supplementary Material compares $F_s(k, t)$ for the FG and CG trajectories at the same temperature and underlying force fields.

**B. Translational Relaxation Time and Temperature Dependence**

Having computed both the diffusion coefficients and the translational relaxation time, a natural direction is to examine the temperature dependency of these dynamical observables. Similar to the Adam-Gibbs theory in liquid dynamics,[169] the following temperature dependency has been suggested by Vogel-Fulcher-Tamman, commonly known as the Vogel-Fulcher-Tamman law[170-172] that links the common dynamical observables $X$ such as diffusion coefficients, shear viscosities, and relaxation times to temperature $T$

$$X = X^0 \exp\left(\frac{A}{T - T_0}\right),$$

(40)

where $X^0, A$ are constants, and $T_0$ is so-called Vogel divergence temperature. Therefore, we first validate if the computed translational relaxation time contains correct dynamical information that satisfies the Vogel-Fulcher-Tamman equation. Consequently, we examine if similar temperature-dependent behaviors are observed in the translational and rotational diffusion coefficients. As introduced in Eq. (40) above, unlike the diffusion coefficients, the translational relaxation timescale is reciprocal to the dynamical properties, resulting in two Vogel-Fulcher-Tamman-like laws

$$\tau = \tau^0 \exp\left(B \frac{T_0}{T - T_0}\right)$$

(41)

$$\frac{1}{D} = \frac{1}{D^0} \exp\left(B \frac{T_0}{T - T_0}\right),$$

(42)

where the unknown parameters $B, T_0, \tau^0$, and $D^0$ are obtained by fitting these relationships. Here, $D^0$ is different from $D_0$, an entropy-free diffusion coefficient from Eq. (1). We note that the Vogel-Fulcher-Tamman law has been widely used for the supercooled regime when the SE relationship breaks down, but we still seek to examine the fidelity of this law at ambient conditions in order to elucidate the temperature-dependence of the computed dynamical properties. Figure 2 depicts three dynamical properties ($\tau, D_{\text{trans}}$, and $D_{\text{rot}}$) computed at temperatures ranging from 280 to 360K. We first confirm that the translational relaxation times obtained by fitting the KWW function to Fig. 1 demonstrate the exponential behavior, as expected from Eq. (41). Furthermore, for four atomistic force fields, we discover that the TIP4P/Ice has the largest relaxation time and decays more rapidly as temperature increases compared to TIP4P/2005, SPC/E, and SPC/Fw. This behavior is consistent with the Vogel-Fulcher-Tamman scaling behavior for diffusion coefficients [see Figs. 2(b) and 2(c)], where the inverse of diffusion coefficients for the TIP4P/Ice model sees the largest changes.

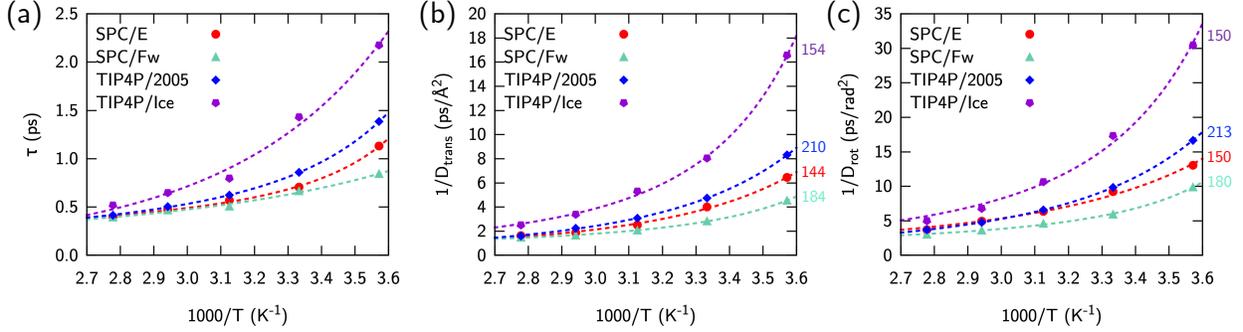

**Figure 2**. The temperature dependence of the various dynamical properties of water: (a) Translational relaxation time, and the inverse of self-diffusion coefficients for (b) translational motion and (c) rotational motion of water. The dotted line in each panel denotes the fitted Vogel-Fulcher-Tamman relation.

In order to validate if both the translational and rotational motions originate from the same molecular nature, we further compare the crossover temperatures $T_0$ obtained from fitting the Vogel-Fulcher-Tamman relation. Surprisingly, we discover that for the same FG force field, both the translational and rotational $T_0$ values are in close agreement. Specifically, SPC/E gives 144 and 150K, SPC/Fw gives 184 and 180K, whereas TIP4P/2005 yields 210 and 213K and TIP4P/Ice yields 154 and 150K (see Table S3 for full parametrized results). In Fig. 2(a), the $T_0$ values derived from $\tau$, which are listed in Table S3, exhibit slightly different values due to the $\alpha$-relaxation but faithfully follow the same Vogel-Fulcher-Tamman relation as temperature changes. Hence, a similar molecular nature underlying the translational and rotational diffusive motions also provides an alternative explanation for why two different motions follow the same excess entropy scaling relationship shown in Eqs. (1) and (15).

We are aware that there are other models designed for explaining the temperature-dependent behavior of dynamical properties, e.g., Speedy-Angell,[173] Bässler,[174] and Krausser-Samwer-Zaccone.[175] Some of these models, including the Vogel-Fulcher-Tamman model used in this section, provides a rather *ad hoc* description. However, we would like to note that the main purpose of this section is not to investigate the microscopic origins of these fitting parameters themselves but to validate whether our simulations correctly exhibit temperature-dependent behavior of translational relaxation time and diffusion coefficients. Particularly, as we are not interested in supercooled conditions, the choice of these models do not significantly impact our temperature-dependent behavior. In turn, our analysis indicates that both the translational relaxation time and the diffusion coefficients can be used interchangeably to describe the dynamical properties of the FG system.

### C. Validity of the SE and SED Relations

We now examine the validity of the SE and SED relationships by utilizing $1/\tau$ as an intermediate variable to approximate $\eta$. In particular, we first validate if the SE and SED relations are well-satisfied for our system. A common practice to examine this is to introduce the *fractional SE and SED relations* defined as

$$D_{\text{trans}} \sim \left(\frac{\tau}{T}\right)^{-\zeta_t},$$

(43)

$$D_{\text{rot}} \sim \left(\frac{\tau}{T}\right)^{-\zeta_r},$$

(44)

where the exponents $\zeta_t$ and $\zeta_r$ indicate how far the system is from the correct SE and SED relations, where $\zeta_t = \zeta_r = 1$. When the SE and SED relations fail at low temperatures (i.e., supercooling regime) below 240K, $\zeta$ is reported to have much lower values near 0.8.[139, 148, 149] Hence, we depicted the log-log plot for the translational and rotational diffusion coefficients versus $\tau/T$ for four FG force fields. Interestingly, we observed that the exponents for both translational and rotational diffusion scalings are 1.1. Because these exponents are near 1.0, it is reasonable to conclude that the SE and SED relations hold in our systems. This is somewhat expected since we choose the temperature ranges to be within the liquid phase range (280–360K) where there is no sign of supercooling. Given the recent success in elucidating the various anomalies associated with the supercooling using BUMPer,[119] an interesting future direction from this analysis would be to study the fractional SE and SED relations for the FG water and BUMPer models in the supercooled regime.

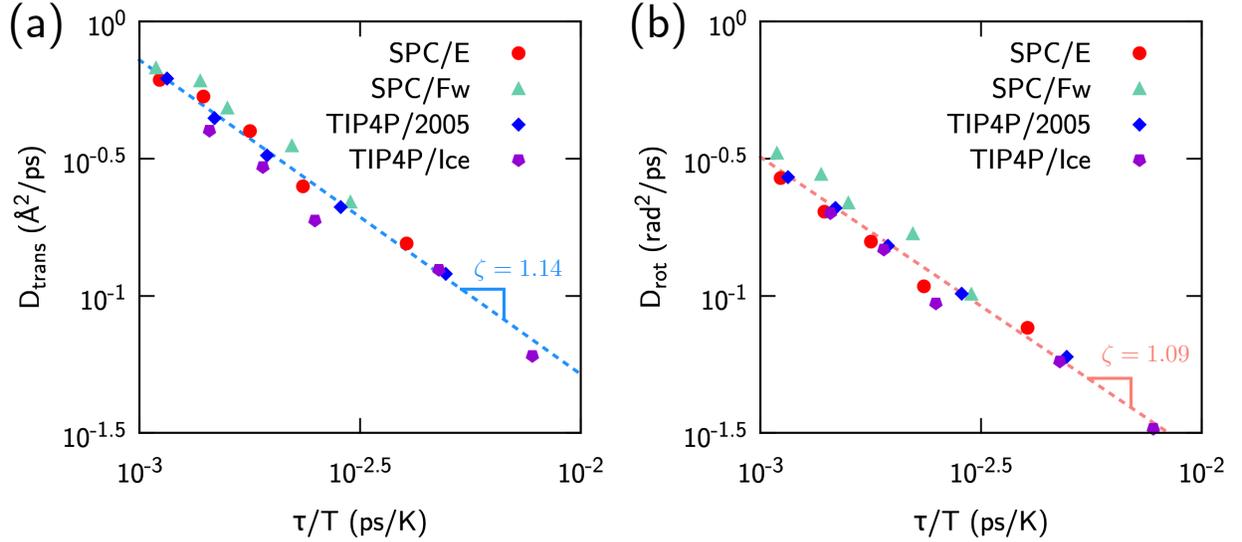

**Figure 3.** Assessment of the SE and SED relations by plotting the translational and rotational diffusion coefficients versus the structural relaxation time divided by the temperature $\tau/T$ for four FG force fields: SPC/E (red circle), SPC/Fw (green triangle), TIP4P/2005 (blue diamond), TIP4P/Ice (purple pentagon). (a) SE relation from the translational diffusion coefficients at different $\tau/T$ and (b) SED relation from rotational diffusion coefficients at different $\tau/T$. The $\tau$ values are from Table 1. The dashed lines (blue: translation, red: rotation) represents the fitted fractional SE and SED relations, $D_{\text{trans}} \sim (\tau/T)^{-\zeta_t}$ and $D_{\text{rot}} \sim (\tau/T)^{-\zeta_r}$, where both exponents $\zeta_t$ and $\zeta_r$ are near 1.1.

It is also worth mentioning that the scaling behavior depicted in Fig. 3 does not explicitly depend on the type of FG force fields. We believe that the universality of the scaling relationship between $D$ and $\tau/T$ is satisfied because the peculiarities and differences among various FG force fields are already encoded in diffusion coefficients and translational relaxation times. Altogether, Fig. 3 confirms that we can utilize the (correct) SE and SED relations to describe the diffusion coefficients for our systems.

Then, the next step is to check the temperature dependence of the SE and SED relations.[117] In doing this, we alternatively examine the temperature dependence of $D_{\text{trans}}\tau/T$ and $D_{\text{rot}}\tau/T$ (known as "Stokes-Einstein ratio"), as shown in Fig. 4. By normalizing $D\tau/T$ at the highest temperatures (360K), we observe that this ratio slightly decreases as with decreasing temperature, but it does not significantly deviate from the reference value expected from the SE and SED relations. We note that the violation of the SE and SED relations at lower temperatures for water are often associated with a magnitude of $D\tau/T$ around 1.5 – 4 for translational diffusion[139, 148, 149] and 1.5 – 8 for rotational diffusion,[157] which is far beyond the values presented in Fig. 4. Finally, based on Figs. 3 and 4, we conclude that the SE and SED relations faithfully hold in our system at the chosen temperature ranges. Combined together, *the next key step is to combine the SE/SED relations with the excess entropy scaling formalism.*

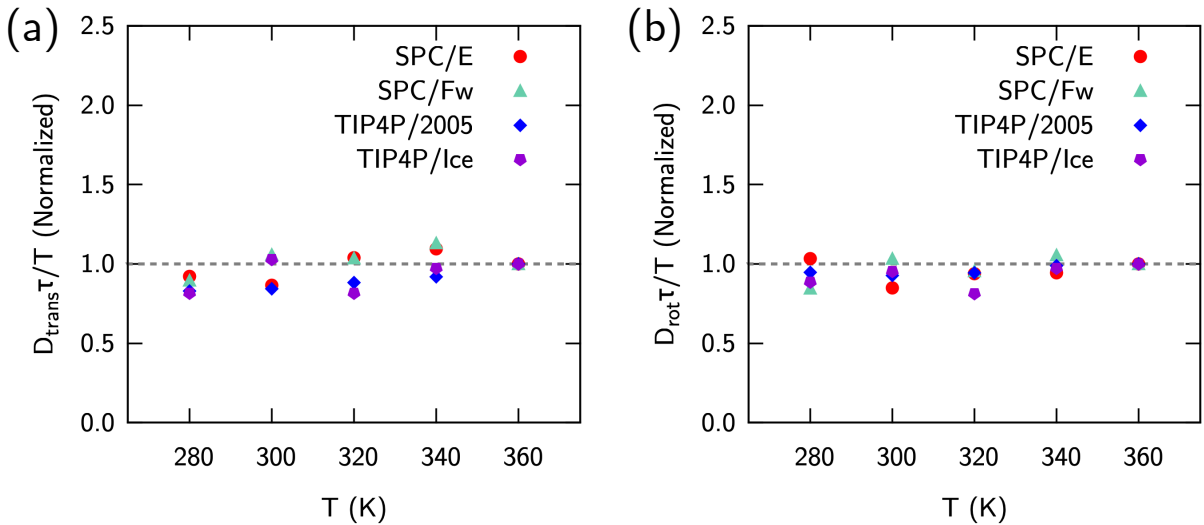

**Figure 4.** Assessment of the SE and SED relations at different (ambient) temperatures ranging from 280 to 360K for four FG force fields: SPC/E (red circle), SPC/Fw (green triangle), TIP4P/2005 (blue diamond), TIP4P/Ice (purple pentagon). (a) SE relationship was examined by computing $D_{\text{trans}}\tau/T$, (b) SED relationship was examined by computing $D_{\text{rot}}\tau/T$, where the plotted data were scaled by their values at $T$=360K.

**D. Determining Hydrodynamic Radius and Reduced Scaling Relationship**

Having confirmed the SE and SED relations using the translational relaxation time, the last remaining variable to estimate in Eqs. (38a) and (38b) is the hydrodynamic radius $R_h$. In experimental settings, $R_h$ can be determined using gel (permeation or filtration) chromatography[176] or NMR diffusion spectroscopy.[177] However, estimating $R_h$ for non-spherical and complex molecules in theory is more ambiguous and challenging.[161, 162, 178] In this direction, an alternative SE relationship could be derived that does not explicitly involve the hydrodynamic radius by reformulating the relation based on reduced (dimensionless) properties[179] and was recently validated for water.[160] Nevertheless, since our main objective of this paper is to be able to indirectly estimate the complex hydrodynamic radius from the hard sphere description, in this subsection we introduce a formal approach that can potentially estimate the hydrodynamic radius of our system, which deviates from a perfect sphere.

To account for the non-spherical nature of molecules, an ellipsoidal approximation of the molecule would provide more flexibility and expressiveness. Formally, a rigorous method to determine the hydrodynamic radius of an ellipsoid was proposed by Perrin under the continuum hydrodynamic case, where the SE and SED relations were initially derived, i.e., a colloidal solute in a continuum medium.[180, 181] In such cases, $R_h$ can be estimated by decomposing the exerted friction along three different inertial axes as follows:

$$R_h = \frac{k_B T}{6\pi\eta D} = \frac{1}{6\pi\eta}\left(\frac{3}{f_1^{-1} + f_2^{-1} + f_3^{-1}}\right)$$

(45)

where $f_i$ ($i \in [1,3]$) denotes the friction coefficients along the three different inertial axes of the particles. Perrin then demonstrated that a friction component along the inertial axes takes the form[180]

$$f_i = \frac{16\pi\eta}{\Pi + a_i^2 P_i}.$$

(46)

Here, $a_i$ are dimensionless shape factors that account for the deviations from spherical symmetry along each principal axis of the ellipsoid, and $P_i$ and $\Pi$ are elliptical integrals to account for the anisotropy (acentricity) defined as follows:

$$P_i := \int_0^\infty \frac{ds}{(s + r_i^2)\sqrt{(s + r_1^2)(s + r_2^2)(s + r_3^2)}},$$

(47)

$$\Pi := \int_0^\infty \frac{ds}{\sqrt{(s + r_1^2)(s + r_2^2)(s + r_3^2)}}.$$

(48)

In Eqs. (47) and (48), $s$ is an integration variable with the dimension of length squared, and $r_i$ are the semi-axes of the ellipsoidal particles, with $r_1$, $r_2$, and $r_3$ specifically referring to the radii along the three principal axes of the ellipsoid. These definitions yield dimensions of length$^{-3}$ for $P_i$ and length$^{-1}$ for $\Pi$, respectively.

Since the diffusion observed in the single-site CG systems shows isotropically averaged behavior,[181] $f_i^{-1}$ in Eq. (45) can be simplified, resulting in the diffusion coefficient of the form

$$D = \frac{k_B T}{4\pi\eta}\Pi.$$

(49)

While the presence of the $\Pi$ term in Eq. (49) reflects the non-spherical nature of the friction due to the intrinsic topology of FG systems,[182] it should be noted that a formal derivation of $\Pi$ is only valid for colloidal Brownian motion under a continuum hydrodynamic description. Yet, inspired by the practical evaluation of $\Pi$ done in hydrated proteins,[183, 184] a reasonable estimation of the $\Pi$ integral can be made by assuming that the elliptical boundaries given by $r_1$, $r_2$, and $r_3$ correspond to the shape of the water molecule.

Once the molecular-level elliptical integral $\Pi$ is available, we can now apply the excess entropy scaling relationship to Eq. (49)

$$D^* = D \frac{\rho^{\frac{1}{3}}}{\left(\frac{k_B T}{m}\right)^{\frac{1}{2}}} = \frac{\rho^{\frac{1}{3}}\sqrt{mk_B T}}{4\pi\eta}\Pi.$$

(50)

We emphasize here again that the dimensionless diffusion $D^*$ in Eq. (50) follows $D^* = D_0 \exp(\alpha s_{ex})$. In order to further elucidate the molecular nature underlying $D_0$, we simplify other variables appearing in Eq. (50) by introducing the viscosity scaling relationship. From Andrade's theory, the dimensionless viscosity can be represented as[185]

$$\eta^* = \eta \cdot \frac{\rho^{-\frac{2}{3}}}{\sqrt{mk_B T}} = \eta_0 \exp(\alpha_\eta s_{ex}).$$

(51)

Hence, $\eta$ can be expressed as

$$\eta = \rho^{\frac{2}{3}}\sqrt{mk_B T}\,\eta_0 \exp(-\alpha_\eta s_{ex}).$$

(52)

We now substitute $\eta$ in Eq. (50) with Eq. (52) to further simplify:

$$D^* = \frac{\Pi \rho^{-\frac{1}{3}}}{4\pi\eta_0}\exp(-\alpha_\eta s_{ex}).$$

(53)

Since $D^* = D_0 \exp(\alpha s_{ex})$, in order to satisfy these two scaling relationships for many different $s_{ex}$ values, the unique solution is if and only if the scaling exponents for both diffusion and viscosity satisfies $\alpha_\eta = -\alpha$ with the entropy-free property:

$$D_0 = \frac{\Pi \rho^{-\frac{1}{3}}}{4\pi\eta_0}.$$

(54)

Equation (54) is the main result of this subsection, as it directly relates the scaling coefficients from different dynamical properties ($D_0$ and $\eta_0$) with the system property $\rho$ and intrinsic molecular property $\Pi$. To note, $\Pi$ can, in principle, be estimated from the molecular-level approximation, as demonstrated in the Supplemental Material. We also note that the skew-symmetric nature established in Eq. (53) between the diffusion and viscosity scaling relationship ($\alpha_\eta = -\alpha_D$) was originally reported by Rosenfeld[186] and confirmed by other follow-up studies.[187-193] Our theory expressed in Eqs. (50)-(54) further proves the universal excess entropy scaling for both diffusion and viscosity with the correct skew-symmetric behavior, which is believed to facilitate the correct scaling behavior for viscosity that has been recently reported in the field[194-198] and also expand our understanding on the correct physical description of the SE relations when applying it to the excess entropy scaling.[199-201] As future work, we plan to accurately determine the shear viscosity in the FG and CG systems to uncover $\eta_0$ values and rigorously bridge between $D_0$ and $\eta_0$ along with $\Pi$.

**E. How to Estimate Hydrodynamic Radius: SE/SED approach and perturbation approach**

Starting from the definition of the hydrodynamic radius, we showed that the SE relation implies the universal excess entropy scaling for dynamical properties. However, an estimation for the elliptical integral Π is built upon numerous approximations, and thus it might contain numerical errors when directly plugged into the SE and SED relations.

On the other hand, a relatively less complicated and more direct approach to compute the hydrodynamic radius is possible by utilizing the coupling between the SE and SED relations. Assuming that the hydrodynamic radii in Eqs. (38a) and (38b) are identical, the hydrodynamic radius can be alternatively expressed using the ratio between $D_{\text{trans}}$ and $D_{\text{rot}}$

$$R_h \approx R_{\text{SE-SED}} = \sqrt{\frac{3}{4}\frac{D_{\text{trans}}}{D_{\text{rot}}}}.$$

(55)

Since we already confirmed that the SE and SED relations hold in our case, we can directly validate this assumption by checking if the ratio $\sqrt{(3D_{\text{trans}})/(4D_{\text{rot}})}$ is constant at different temperatures. Figure 5(a) shows the ratio between the diffusion coefficients $D_{\text{rot}}/D_{\text{trans}}$ as a function of temperature.

The ratio of the translational and rotational diffusion coefficients is essentially a measure of the decoupling between the two motions. It has been widely studied that for glass-forming liquids[137, 202, 203] and water,[157, 204, 205] the translation-rotation decoupling is observed in the supercooling regime where the SE and SED relations break down. In the SE/SED breakdown case, the $D_{\text{rot}}/D_{\text{trans}}$ ratio is known to rapidly decrease as temperature lowers[157] because the SED relation breaks down quicker than the SE relation.[139] However, in our case, we note that the diffusion ratio does not greatly change with decreasing temperature [Fig. 5(a)], with an average of 0.428 rad$^2$/Å$^2$ for SPC/E, 0.467 rad$^2$/Å$^2$ for SPC/Fw, 0.471 rad$^2$/Å$^2$ for TIP4P/2005, and 0.502 rad$^2$/Å$^2$ for TIP4P/Ice. Even though these ratios fluctuate slightly with a standard deviation of 0.038 rad$^2$/Å$^2$, $D_{\text{rot}}/D_{\text{trans}}$ does not show the breakdown behavior. Thus, we assert this fluctuation is negligible, and the hydrodynamic radius can be estimated by the ratio between the two diffusion coefficients.

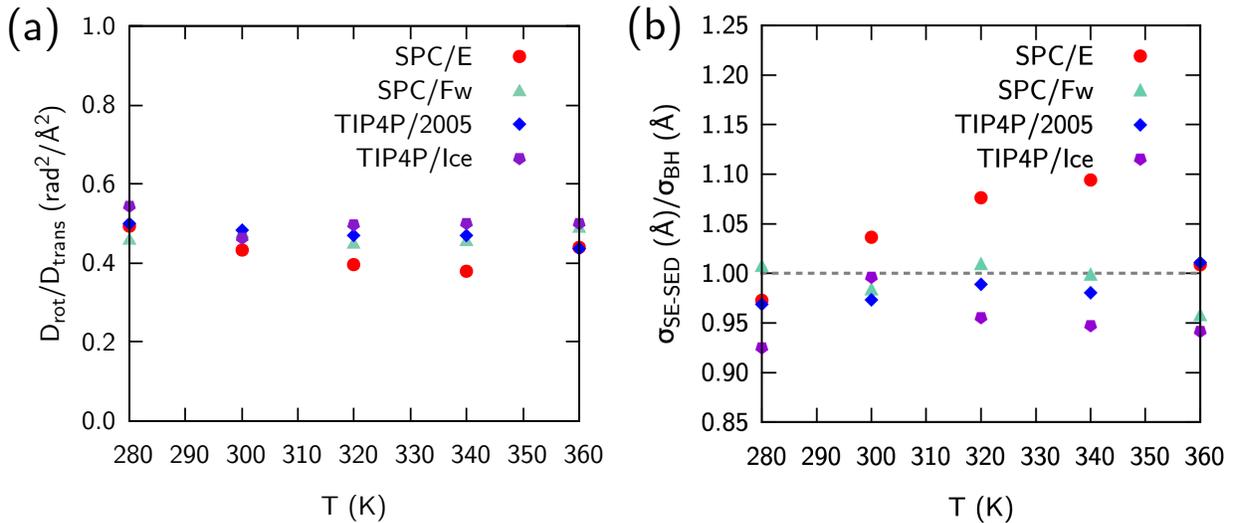

**Figure 5.** Validation and estimation of the hydrodynamic radius using two approaches for four FG force fields: SPC/E (red circle), SPC/Fw (green triangle), TIP4P/2005 (blue diamond), TIP4P/Ice (purple

pentagon). (a) Validation of the converged hydrodynamic radius using the SE and SED relations by computing the ratio of the rotational and translational diffusion coefficients $D_{\text{rot}}/D_{\text{trans}}$ as a function of temperature. (b) The ratio of the hydrodynamic diameter derived from the SE and SED relations $\sigma_{\text{SE-SED}} \approx 2R_h$ [Eq. (55)] and the Barker-Henderson diameter $\sigma_{\text{BH}}$.

Having established the hydrodynamic radius, the next question we want to address is how the hydrodynamic radius is related to the excess entropy scaling. We now provide a new understanding of the hydrodynamic radius that is physically consistent with the excess entropy scaling formalism. As extensively discussed in Paper II of this series,[35] the CG diffusion process can be described by a hard sphere diffusion process. Specifically, in terms of the excess entropy scaling, the CG diffusion can be alternatively expressed as

$$D_{\text{CG}}^* = D_0^{\text{CG}} \exp(\alpha^{\text{CG}} s_{ex}^{\text{CG}}) = D_0^{\text{CG}}(\eta) \exp(\alpha^{\text{CG}} s_{ex}^{\text{CG}}),$$

(56)

where a statistical mechanical link can be established between $D_0^{\text{CG}}$ and the hard sphere nature of the CG system represented as a packing density $\varphi$ or the effective hard sphere diameter (EHSD) $\sigma_{\text{EHSD}}$. From the definition of the packing density, $\varphi$ is related to the EHSD $\sigma_{\text{EHSD}}$ via

$$\varphi = \frac{\pi}{6} \sigma_{\text{EHSD}}^3 \rho,$$

(57)

where $\rho$ is the number density of the system.

From the molecular CG system, $\sigma_{\text{EHSD}}$ can be estimated based on the CG interaction profile. The Barker-Henderson perturbation theory[206, 207] suggests that $\sigma_{\text{EHSD}}$ can be reasonably approximated as the "repulsive" regimes at short ranges by calculating the Barker-Henderson diameter $\sigma_{\text{BH}}$

$$\sigma_{\text{BH}} = \int_0^{R_0} [1 - \exp(-\beta U(R))] \cdot dR,$$

(58)

where $R_0$ is the minimum distance where the interaction vanishes, i.e., $U(R_0) = 0$. Therefore, a physical picture behind the Barker-Henderson diameter is that it is the effective "size" of the system based on its repulsive core. Our approach shares a similar physics with the early work by Bocquet et al. where they construed the hydrodynamic radius as the radius of the excluded volume.[208, 209] Yet, Eq. (58) is a generalized description for the reduced CG representation. Since Eq. (58) accounts for the weak repulsive contribution at larger distances [as long as $U(R) > 0$], we expect that the $\sigma_{\text{BH}}$ might provide a slightly larger estimate observed from the interaction profile than the molecule's actual size. Nevertheless, as long as the fluid molecule experiences the dissipative and fluctuation forces that can be described by the hydrodynamic process, we envisage that the *Barker-Henderson radius* $R_{\text{BH}} = \sigma_{\text{BH}}/2$ to be in close agreement with $R_h$. By comparing the Barker-Henderson diameters to the hydrodynamic diameter estimated from the atomistic simulations using Eq. (55), $\sigma_{\text{SE-SED}} = 2R_{\text{SE-SED}}$, Fig. 5(b) confirms that one could utilize the CG description of molecules to assess a relatively challenging to evaluate hydrodynamic radius over a wide range of temperatures.

**F. New Theory for Translational-Rotational Diffusion Coefficients based on the SE/SED Relations**

For more general cases, Fig. 5(b) indicates that the hydrodynamic diameter can be reasonably approximated from the EHSD of the CG system. Therefore, we can further relate the excess entropy scaling formalisms for the translational and rotational diffusion in terms of the SE and SED relations when the following two assumptions are valid, i.e., $\sigma_{SE-SED} \approx \sigma_{EHSD}$, and there is no SE or SED breakdown.

Here, the translational diffusion for the FG system given by the SE relation obeys the following excess entropy scaling relationship:

$$D^*_{trans} = \frac{k_B T}{6\pi\eta R_{CG}} \cdot \left(\frac{\rho^{1/3}}{\sqrt{k_B T/m}}\right) = D_0^{FG} \exp(\alpha^{trans} s_{ex}^{FG}),$$

(59)

where we denote the (unified) CG radius as $R_{CG}$ that is consistent with the SE/SED and EHSD description

$$R_{CG} = \frac{1}{2}\sigma_{CG} \approx \frac{1}{2}\sigma_{SE-SED} \approx \frac{1}{2}\sigma_{EHSD}.$$

(60)

Similarly, the projected rotational diffusion can be expressed using the excess entropy scaling

$$D^*_{t-rot} = \frac{2}{3} D_{rot} \cdot R_{CG}^2 \left(\frac{\rho^{1/3}}{\sqrt{k_B T/m}}\right) = D_0^{t-rot} \exp(\alpha^{t-rot} s_{ex}^{FG}).$$

(61)

Note that the constant 2/3 in Eq. (61) comes from the normalization factor during a transformation of the rotational diffusion to the translational diffusion [the $4t$ factor in $D_{rot}$ in Eq. (32) needs to be corrected to $6t$ in Eq. (31)]. Introducing the SED relationship [Eq. (38b)] to Eq. (61) gives

$$D^*_{t-rot} = \frac{k_B T}{12\pi\eta R_{CG}^3} \cdot R_{CG}^2 \left(\frac{\rho^{1/3}}{\sqrt{k_B T/m}}\right) = D_0^{t-rot} \exp(\alpha^{t-rot} s_{ex}^{FG}).$$

(62)

From Paper III, we know that the exponent for the scaling relationships of translational and projected rotational diffusion are nearly identical ($\alpha^{trans} \approx \alpha^{t-rot}$), and thus the entropy parts of the diffusion relationship are canceled by dividing Eq. (59) by Eq. (62), resulting in

$$2 = \frac{D_0^{FG}}{D_0^{t-rot}}.$$

(63)

An interesting observation is that the left-hand side of Eq. (63) no longer has $R_{CG}$ since $R_{CG}$ is canceled by projecting the rotational diffusion to the arc displacement, resulting in a dimensionless value. Also, Eq. (63) states that the entropy-free diffusion at the FG (translational) level is directly related to its translational-rotational component. This direct relationship further suggests that one does not have to explicitly calculate the rotational diffusion at the FG level (and then project them onto the translational basis), as long as the SE and SED relationships are valid.

Our next question is how this relationship impacts the full dynamic consistency between the FG diffusion and CG diffusion. Recall that the FG and CG diffusion coefficients are linked by the following relationship from Eq. (18),

$$D_0^{FG} = D_0^{CG} D_0^{t-rot} A_D.$$

(64)

Then, Eq. (63) immediately suggests that Eq. (64) can be further reduced to

$$D_0^{CG} = \frac{2}{A_D}.$$
(65)

Equation (65) implies that the entropy-free diffusion coefficient of the single-site CG model is only a function of the roughness parameter, $A_D$. In Eq. (17), we introduced an approach from Ruckenstein and Liu[50] where they tried to interpret $A_D$ as a function of the non-sphericity. Then, combining Eqs. (17) and (65) gives

$$D_0^{CG} = \frac{2}{0.9673 - 0.2527\omega - 0.70\omega^2}.$$
(66)

Now, the physical meaning of Eq. (66) becomes clearer: the entropy-free diffusion coefficient of the CG system (thus translational motions) can be determined solely by the non-sphericity of the CG molecule as long as the SE and SED relationships hold. This physical picture can be viewed in a similar manner as the main result in Paper II, where we introduced the hard sphere mapping theory to determine $D_0^{CG}$ as a function of the packing density $\varphi$.[35] Recall that the analytical expression for $D_0^{CG}$ using the Carnahan-Starling EOS[91] is reduced to Eq. (13):

$$D_0^{CG}(\varphi) = D_{0,CS}^{HS} = \frac{\pi^{\frac{1}{6}}}{48} 6^{\frac{4}{3}} \frac{(1-\varphi)^3}{\eta^{\frac{2}{3}}(2-\varphi)} \exp\left[\frac{4\varphi - 3\varphi^2}{(1-\varphi)^2}\right].$$
(67)

Figure 6 summarizes the two different approaches for calculating the $D_0^{CG}$ values.

Some differences between these two approaches should be noted. First, by definition, the hard sphere mapping approach [Eq. (13)] represented by Fig. 6(a) is only valid for the hard sphere description. If the system deviates from the hard sphere description, the systematic methodology for determining the EHSD, such as the Barker-Henderson perturbation theory and fluctuation matching, is no longer valid and results in an unphysical packing density.

However, the second approach delineated in Fig. 6(b) is not limited by the same issue. As long as the target system at the FG resolution exhibits both translational and rotational motions, there must be a momentum exchange upon collision between the system particles. This gives rise to a non-unity value for $A_D$. At the same time, if the system is at ambient conditions such that the target system satisfies the SE and SED relations, then both the translational and rotational motions are coupled to each other. By combining these two descriptions, we arrive at Eq. (65), where no further approximation is needed. The only approximation used in Eq. (66) is the functional form of $A_D$ in terms of $\omega$ by Ruckenstein and Liu,[50] which might be improved in future work.

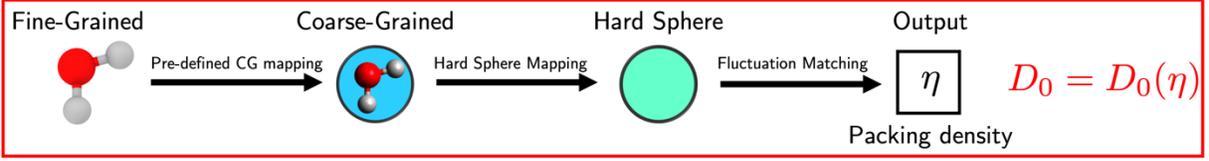
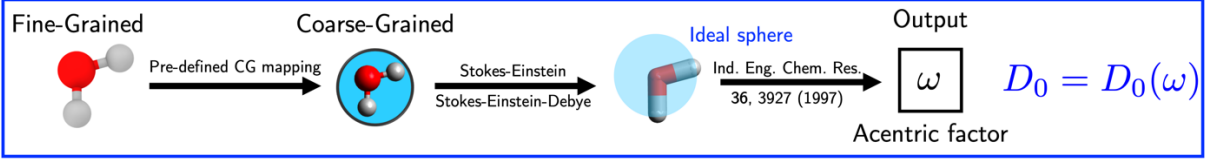

**Figure 6**. Two different approaches to understand the entropy-free diffusion coefficient $D_0$ from the CG model: (a) Hard sphere approach from Paper II giving Eq. (13),[35] (b) SE/SED approach from this work giving Eqs. (66) and (67) with use of Ref. 50. The approach illustrated in (b) provides a much wider and extended view on $D_0$ than (a).

Recall that our original purpose in introducing the SE and SED relation was to deduce the effective rotational diffusion without actually calculating it at the FG level. By assuming that the hydrodynamic radius is the thermodynamic radius from the Barker-Henderson criteria, we can now indirectly estimate the rotational diffusion $\widehat{D}_{\text{rot}}$ using the computed translational diffusion coefficients and estimate the $\widehat{D}_0^{\text{t-rot}}$. Note that the predicted variable is denoted by a hat notation. In Fig. 7, we assess the performance of the SE and SED relations by comparing the rotational diffusion coefficient $D_{\text{rot}}$ from the FG simulations to the predicted diffusion coefficient $\widehat{D}_{\text{rot}}$

$$\widehat{D}_{\text{rot}} = \frac{3}{4} D_{\text{trans}} \cdot \left(\frac{1}{R_{\text{BH}}}\right)^2 = \frac{3 D_{\text{trans}}}{\left[\int_0^{R_0} [1 - \exp(-\beta U(R))] \cdot dR\right]^2}. \tag{68}$$

For clarity, we plotted the results obtained from Eq. (68) against the Vogel-Fulcher-Tamman relation fitted from Fig. 2(c). As consistent with Fig. 6, we also observe that Fig. 7 confirms the fidelity of Eq. (68), indicating that the rotational motions at the FG resolution do not need to be explicitly calculated using complicated methods such as imposing polarization vectors or calculating the rotational correlation functions. Instead, under ambient conditions, the SE and SED relations not only act as an intermediate variable to predict the collective rotational motions embedded in the system at the hydrodynamic level but also confirm our initial findings regarding the rotational scaling relationship. In summary, this work demonstrates how these relations can widen our understanding of liquid dynamics at both the FG and the CG levels.

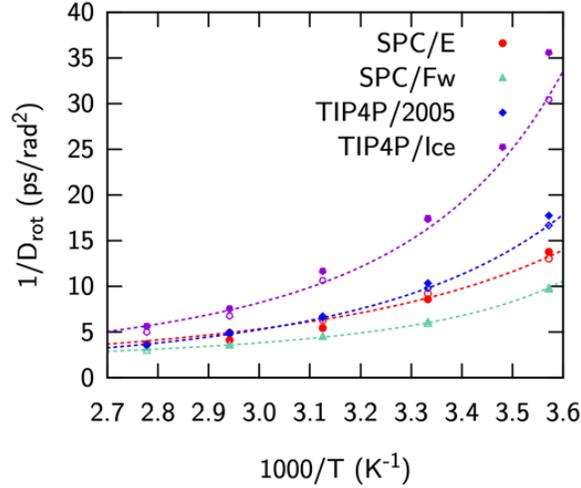

**Figure 7.** Assessing the fidelity of diffusion prediction based on the SE and SED relations for four FG force fields: SPC/E (red circle), SPC/Fw (green triangle), TIP4P/2005 (blue diamond), TIP4P/Ice (purple pentagon). Predicted diffusion coefficients $\widehat{D}_{\text{rot}}$ using Eq. (68) are shown as filled points, whereas the actual values from Fig. 2(c) are marked as empty points. The dotted lines are from fitting the FG rotational diffusion coefficients to a Vogel-Fulcher-Tamman relation.

## IV. CONCLUSIONS

This paper builds upon our earlier work on the correspondence between FG and CG dynamics[34-36] based on the hydrodynamic description. In particular, we introduce the SE and SED relations to the CG system in order to extend the applicability of the currently established framework. In the previous papers of this series, we had shown that information about the missing rotational diffusion may play an important role in understanding the differences in the diffusion coefficients of FG and CG liquids under Hamiltonian mechanics, in the limit of a "one bead" CG mapping. The earlier Paper III suggested that the rotational motions at the FG resolution, which are integrated out during the coarse-graining process, are responsible for this deviation.[36] However, it was still ambiguous how to extract this missing information at the reduced resolution. This work fills this gap by combining the commonly used SE and SED relations with the excess entropy scaling relationship.

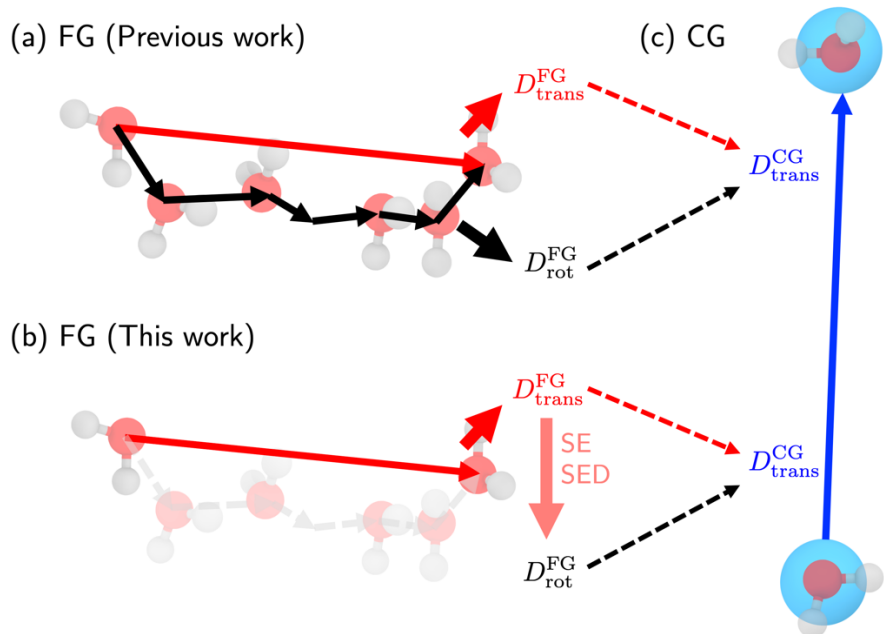

**Figure 8.** Role of the SE and SED relations in elucidating the FG [panels (a) and (b)] and CG diffusion [panel (c)]. (a) In terms of the excess entropy scaling, Paper III suggested that the original FG diffusion (red line) $D_0^{FG}$ can be recapitulated by combining the CG diffusion (blue line) $D_0^{CG}$ and the missing rotational diffusion (black line) at the finer resolution $D_0^{rot}$, where $D_{rot}$ was computed by integrating the differential of the angular displacement vector. (b) Alternatively, this work suggests that there is no need to explicitly calculate the $D_{rot}^{FG}$, as this value can be deduced by employing the SE and SED relations to the computed $D_{trans}^{FG}$ value.

Figure 8 summarizes the essential findings from this work. As both the SE and SED relations are a function of viscosity, a property that is highly challenging to correctly represent in the CG system due to the representability issue, we propose using the translational relaxation time from the incoherent intermediate scattering function as an intermediate variable to approximate the viscosity. We confirmed that the SE and SED relations hold for our system at ambient temperatures by checking the fractional SE and SED relations and also by computing the SE and SED ratios. With this in mind, we provided an alternative approach to determine the hydrodynamic radius of the system using the EHSD given by the classical perturbation theory. In turn, we corroborated excellent agreement between the Barker-Henderson radius and the hydrodynamic radius from the SE and SED ratios. On the other hand, from approximating the elliptical integrals from the inertial frictions, we derived that the excess entropy scaling for viscosity is universal to the diffusion scaling relationship but with an opposite sign in the entropy term. By further linking our findings to the entropy scaling relationship, an interesting connection between the entropy-free diffusion coefficient and the roughness parameter is established: we obtained an extended description of the entropy-free diffusion coefficient from our original understanding of the hard sphere nature of the CG diffusion process.

We conclude this paper by discussing a few promising and important directions for future work. These directions involve (1) generalizations of the proposed framework and (2) applications to more complex systems beyond liquids. First, throughout this series of papers, we have only considered the single-site CG resolution. As a result, decoupling between translational and rotational motions was relatively straightforward for the single-site (one bead) CG resolution because the rotational motions disappear at this resolution. Therefore, the next step in terms of applicability would be to extend the developed approaches to CG resolutions other than the single-site, where these two motions are no longer decoupled at the CG level. This also requires a systematic methodology to determine the exact modal entropy and its ideal gas contribution at nontrivial resolutions.

Relatedly, another inherent limitation of the scaling relationships employed in this series is that they are only applicable to isotropic dynamical properties. This constrains the detailed description of the system dynamics, as viscosity should be treated as a stress tensor, and diffusivity should be anisotropic in principle. Importantly, anisotropy is pronounced in rotational diffusion, known as the anisotropic rotational diffusion tensor. While this anisotropy effect is relatively minor in water,[210] a faithful generalization to anisotropic diffusivity is crucial for understanding the dynamics of complex molecules, e.g., high-density protein solutions.[211, 212] However, since the semi-empirical nature of the scaling relationship has only been established for scalar dynamical properties,[85] there is no guidance on how this *ad hoc* scaling relationship could be generalized for tensorial representation, nor is there a conceptual demonstration of how this would work in this case, highlighting the need for future work to address these ambiguities. While there are currently no microscopic grounds for the anisotropic generalization, we would like to note that the transport process in confined systems (e.g., nanoporous system) would be an ideal starting point for testing the validity of tensorial scaling relationships. Generally, transport diffusivity is described via Fick's law, but the confined geometry of nanoporous systems gives rise to anisotropic diffusivity. Reference 213 showed that the scalar diffusivity, obtained by averaging the three diagonal elements of the diffusion tensor, $D_t = (D_{t,xx} + D_{t,yy} + D_{t,zz})/3$, obeys excess entropy scaling under the low-density approximation, which assumes that self-diffusivity and transport diffusivity are identical.[214] Therefore, generalizing the current framework to understand how tensorial scaling averages into scalar scaling would significantly enhance its applicability to describe complex molecules.

In a related fashion, a systematic generalization of the current methodology is possible by broadening the scope of this research to study different systems beyond simple liquids. Even though most of the work on the SE and SED relations has been primarily focused on liquids (especially supercooled liquids), it has been shown that these relations also hold for the diffusion of proteins, macromolecules,[215-218] and nano-particles.[219] Therefore, we believe that it may be feasible to extend the present understanding and our theory to more complex systems. One such example would be polymeric systems, where the excess entropy scaling relationship at the atomistic scale has been already reported.[220-224] For relatively simple polymers, we could effectively estimate the hydrodynamic radius using the Kirkwood formula[225-227] or the Kirkwood–Riseman hydrodynamic equation,[227, 228] which can be directly utilized for the SE and SED relations. Additionally, another related avenue would be to extend the reported analyses to glass-forming liquids at the CG level, e.g., ortho-terphenyl[229] or to other experimentally relevant materials.[230, 231] This broader applicability would facilitate the integration of excess entropy scaling into the exploration of slow dynamics near the glass transition. While this future work poses significant challenges due to the state point-dependent nature of CG interactions and the representability problem in evaluating various CG properties, we believe that the systematic CG methodology presented in this work lays the groundwork for advancing this direction. We hope to report a variety of new directions and applications in future publications.

**ACKNOWLEDGMENTS**
This material is based upon work supported by the National Science Foundation (NSF grant CHE-2102677). Simulations were performed using computing resources provided by the University of Chicago Research Computing Center (RCC). J.J. thanks the Arnold O. Beckman Postdoctoral Fellowship for funding and academic support.

**DATA AVAILABILITY**
The data that support the findings of this work are available from the corresponding author upon request.

**SUPPLEMENTARY MATERIAL**
See supplementary material for details of the incoherent intermediate scattering function for CG trajectories, a full table containing the fitting parameters used in the figures, and the evaluation of the molecular-level elliptical integral from Perrin's formula.